\documentclass[preprint]{aastex}
\pdfoutput=1
\usepackage{graphicx}
\usepackage{apjfonts}
\usepackage{natbib}
\makeatletter

\begin{document}

\title{A \emph{Spitzer} View of The Giant Molecular Cloud Mon OB1 East/NGC 2264}

\shorttitle{A \emph{Spitzer} View of Mon OB1 East/NGC 2264}

\author{Rapson, Valerie A.\altaffilmark{1}, 
	Pipher, Judith L.\altaffilmark{2},
	 Gutermuth, Robert A.\altaffilmark{3}, 
	 Megeath, S. Thomas \altaffilmark{4}, 
	 Allen, Thomas S.\altaffilmark{4}, 
	  Myers, Philip C.\altaffilmark{5}, 
	  Allen, Lori E.\altaffilmark{6}}

\altaffiltext{1}{School of Physics and Astronomy, Rochester Institute of Technology,  Rochester, NY 14623, USA}
\altaffiltext{2}{Department of Physics and Astronomy, University of Rochester, Rochester, NY 14627, USA}
\altaffiltext{3}{Five College Astronomy Department, Smith College, Northampton, MA 01063, USA}
\altaffiltext{4}{University of Toledo, Ritter Astrophysical Observatory, Department of Physics and Astronomy, Toledo, OH 43606, USA}
\altaffiltext{5}{Harvard-Smithsonian Center for Astrophysics, Cambridge, MA 02138, USA}
\altaffiltext{6}{National Optical Astronomy Observatories, Tucson, AZ 85719, USA}

\email{var5998@rit.edu}

\begin{abstract}
We present \emph{Spitzer} 3.6, 4.5, 5.8, 8.0, and 24 $\mu$m images of
the Mon OB1 East giant molecular cloud, which contains the young star forming
region NGC 2264, as well as more extended star formation. 
 With \emph{Spitzer} data and 2MASS photometry,
we identify and classify young stellar objects (YSOs) with dusty circumstellar disks and/or envelopes in Mon OB1 East
by their infrared-excess emission and study their distribution with
respect to cloud material. We find a correlation between
the local surface density of YSOs and column density of molecular gas as traced
by dust extinction that is roughly described as a power law in these
quantities. NGC 2264 follows a power law index of $\sim$2.7, exhibiting a large YSO surface density for a given gas column density. 
Outside of NGC 2264  where the surface density of YSOs is lower, the power law is shallower and the region exhibits a larger gas column density for a YSO surface density, suggesting the star formation is more recent. In order to measure the fraction of cloud members with circumstellar disks/envelopes, we estimate the number of diskless pre-main sequence stars by statistical removal of background star detections. We find that the disk fraction of the NGC 2264 region is 45\%, while the surrounding more distributed regions show a disk fraction of 19\%. This may be explained by the presence an older, more dispersed population of stars.  In total, the Spitzer observations provide evidence for heterogenous, non-coeval star formation throughout the Mon OB1 cloud. 
\end{abstract}


\maketitle
\section{Introduction}

Recent surveys \citep[e.g.,][]{Allen2012, Gutermuth2011, Evans2009, LadaLada2003, Carpenter2000} show that rich, massive young clusters in giant molecular clouds (GMCs) are relatively rare in the nearest kiloparsec. With the exception of the Orion Nebula Cluster \citep{Megeath2012} and Cep OB3b \citep{Allen2012}, most clusters within 1 kpc distance contain less than 1000 members.  NGC 2264, a young cluster within the Mon OB1 East giant molecular cloud (hereafter Mon OB1 East),  is another exception. This cluster contains up to $\sim$1400 members \citep{Teixeira2012}, and is thus slightly smaller than the Orion Nebula Cluster and Cep OB3, but an outlier amongst young clusters within 1 Kpc of the sun \citep{Gutermuth2011, LadaLada2003, Porras2007}.

\emph{Spitzer} Space Telescope images of these star forming
regions can potentially provide both 3.6-8.0 $\mu$m
and 24 $\mu$m photometry to identify young stellar object (YSO) cluster members with circumstellar disks and/or envelopes through their infrared excess emission. Here, we present mid-infrared photometry of Mon OB1 East, including NGC 2264. Infrared photometry alone cannot distinguish diskless pre-main sequence stars from field stars, since neither exhibit infrared excess, so other techniques must be utilized to determine the total number of cluster members.  Often, a combination of X-ray and H-$\alpha$ observations, along with proper motion studies, are used to determine cluster membership \citep[e.g.][]{Allen2012, Wolk2010, Getman2007,Flaccomio2006,Herbig2006}. For large molecular clouds, these data are often not available for the entire region, as is the case for Mon OB1 East, so statistical estimates of membership must be made. 

The density and distribution of stars throughout GMCs reveal much about the age and evolutionary state of the clouds. Studies of stellar distribution in GMCs \citep[e.g.][]{Carpenter2000, Gutermuth2009} have shown that most clouds have both embedded dense young stellar clusters and more diffusely distributed populations of young stars, and Mon OB1 East is no exception. Here, we examine the correlation between the spatial distribution of YSOs and the molecular cloud column density throughout Mon OB1 East to further understand its evolutionary state, as compared with other massive giant molecular clouds such as Mon R2, Cep OB3 and Orion \citep{Gutermuth2011}.

The distance to Mon OB1 East has been estimated between 700 and 950 pc \citep[][and references therin]{Baxter2009, Dahm2008};  here we adopt 760 pc \citep{Park2000, Sung1997} as this is the most commonly adopted value in recent literature. Portions of the cloud are estimated to be $\sim$5 Myr old \citep{Dahm2005}, but numerous associated molecular outflows \citep[e.g][]{Margulis1988, Wolf-Chase2003} and Herbig-Haro objects \citep[e.g][]{Reipurth2004I} have been identified within this region, suggesting that star formation is still occurring. NGC 2264 contains at least three spectroscopically identified young sub-clusters \citep{Furesz2006}. These include the region around IRS-1 or Allen's source \citep{Allen1972}, a 9.5 M$_{\odot}$ B2 zero-age main sequence star associated with the Cone Nebula;
IRS-2, a protostar associated with the Spokes cluster  \citep{Teixeira2006} that is surrounded by a cluster of protostars and sub-millimeter cores \citep{Wolf-Chase2003,Williams2002}, and the young sub-cluster centered on the O7 giant S Mon. These small clusters have been extensively studied by \citet{Sung1997,Sung2004, Sung2007, Sung2008, Sung2009} and \citet{Teixeira2012}. 

Here we present a survey of the Mon OB1 East GMC.
This combines a survey of the NGC 2264 cluster \citep{Sung2008, Sung2009, Teixeira2006, Teixeira2012} with a new survey of the filamentary cloud north and south of the cluster. We classify YSOs using \emph{Spitzer} aperture photometry and the classification criteria described by \citet{Gutermuth2009} in order to provide a census of the YSOs over the entire Mon OB1 East GMC. In section 2 we discuss the sample and the observations. Section 3 describes the YSO classification and gives a total YSO count
for the region. Calculation of disk fractions for Mon OB1 East
is described in section 4, and compared with other work by \citet{Sung2009} and \citet{Teixeira2012}. The method used to calculate extinction
towards Mon OB1 East is described in section 5. In section 6, we analyze the spatial
distribution of YSOs versus dust extinction maps created
from near-infrared (J,H,K) data and study the correlation between molecular gas mass column density and YSO mass surface density. We find that Mon OB1 East has clusters of sources at various evolutionary stages, and compare our results with those obtained for other nearby star forming GMCs. 

\section{Observations and Data Analysis}
	
We utilize 3.6-8.0 $\mu$m images on Mon OB1 East obtained with
the \emph{Spitzer} Space Telescope Infrared Array Camera \citep[IRAC][]{Fazio2004},
24 $\mu$m images obtained with the Multi-Band Imaging Photometer (MIPS; \citet{Rieke2004}),
along with 1-2.5 $\mu$m NIR data from the 2 Micron All Sky Survey
\citep[2MASS-][]{Skrutskie2006}, to classify YSOs. These YSOs in Mon OB1 East are classified as either protostars or stars with circumstellar disks by their
infrared excess emission above photospheric emission. \emph{Spitzer} data were
gathered as part of two Guaranteed Time Observation programs and one additional program, with
the goal of studying clustered and distributed star formation
throughout Mon OB1 East and comparing the results with those of
other molecular clouds.

Mon OB1 East was observed by \emph{Spitzer} in 2004, 2007, and
2008 as part of the Guaranteed Time Observation programs 37 (IRAC data; PI: G.
Fazio) and 58 (MIPS data; PI: G. Rieke), as well as program 40006 (IRAC+MIPS data; PI: G. Fazio)%
\footnote{IRAC AORkeys: 21814272, 21814528, 21814784, 21815040, 21815296, 21815552,
21815808, 21816064, 3956480, 3956736, 3956992, 3957248.\\
MIPS AORkeys: 21816320, 21816576, 21816832, 21817088, 4317184.
}. In programs 37 and 40006, Mon OB1 East was mapped in all four IRAC bands (3.6, 4.5, 5.8
and 8.0 $\mu$m). Observations were performed in two of the four IRAC
bands at two dither positions each observation using the 12 s IRAC High Dynamic Range
mode with exposure times of 0.4 and 10.4 s. The
total area mosaicked was approximately 3.0$^\circ$ x 1.5$^\circ$. Basic calibrated data
(BCD) reduction was completed with the \emph{Spitzer} Science Center
pipeline version 18.7.0. Further post-pipeline processing was
performed with Cluster Grinder, an IDL software package \citep{Gutermuth2008,Gutermuth2009}. Within the package, automated point-source
aperture photometry was performed via PhotVis (V1.10) using synthetic
apertures of 2\arcsec.4 radius, and background annuli of inner and outer
radii of 2\arcsec.4  and 7\arcsec.2, respectively. These aperture and annuli sizes were chosen as they adequately sample the point spread function and minimize contamination from resolved background emission \citep{Megeath2004}. The zero points used are 19.455, 18.699, 16.498, and 16.892 for 3.6, 4.5, 5.8, 8.0 $\mu$m, respectively, all of which are Vega Standard magnitudes for data in units of DN s-1 corrected to the apertures and sky annulus adopted for this paper \citep{Gutermuth2008}; these numbers are based on the calibration of IRAC in \citet{Reach2005}.

MIPS data were obtained in 2004 and 2008 at the medium scan rate with
an exposure time of 3.67 s. Twenty eight total scans of either 0.75$^\circ$
or 0.5$^\circ$ length and 160\arcmin ~offsets were
used. Total integration time for the 24 $\mu$m data was 80 s. These data were processed using the \emph{Spitzer} Science Center
pipeline version 16.1.0, and further reduced using Cluster Grinder
and PhotVis. A synthetic aperture of 7\arcsec.6, background annuli
with inner and outer radii of 7\arcsec.6 and 17\arcsec.8, respectively, and an aperture corrected zero point of 14.6 mag \citep{Gutermuth2008} were used for the 24 $\mu$m waveband.

Portions of the data from programs 37 and 58, specifically that within the NGC 2264 region,  have been published in \citet{Teixeira2006, Teixeira2012} and \citet{Sung2009}. Here, we conduct an independent analysis of all data associated with Mon OB1 East.

\section{Identifying Young Stellar Objects}

\subsection{Identification Methods}

\begin{figure}
\includegraphics[scale=0.5]{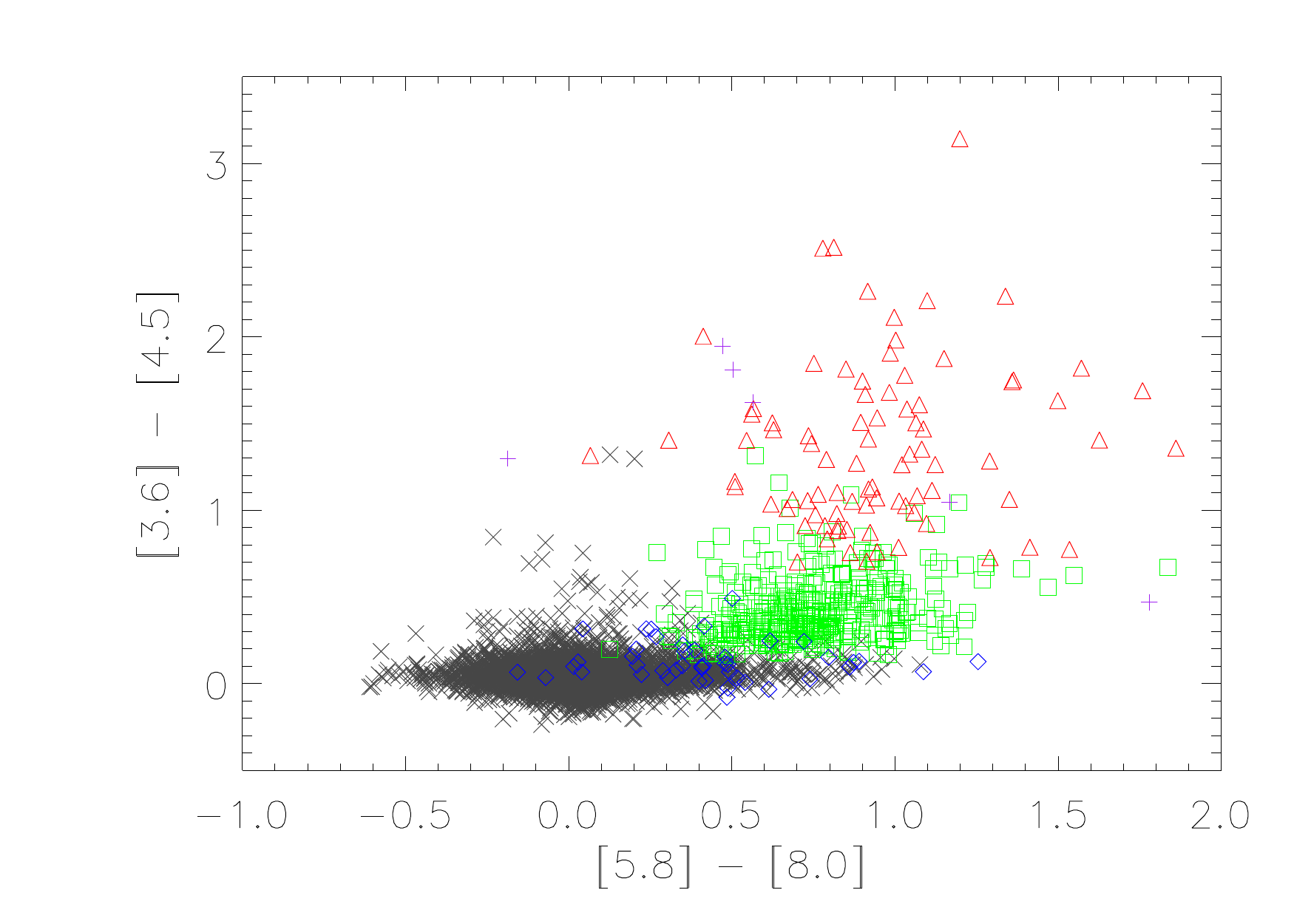} 
\includegraphics[scale=0.5]{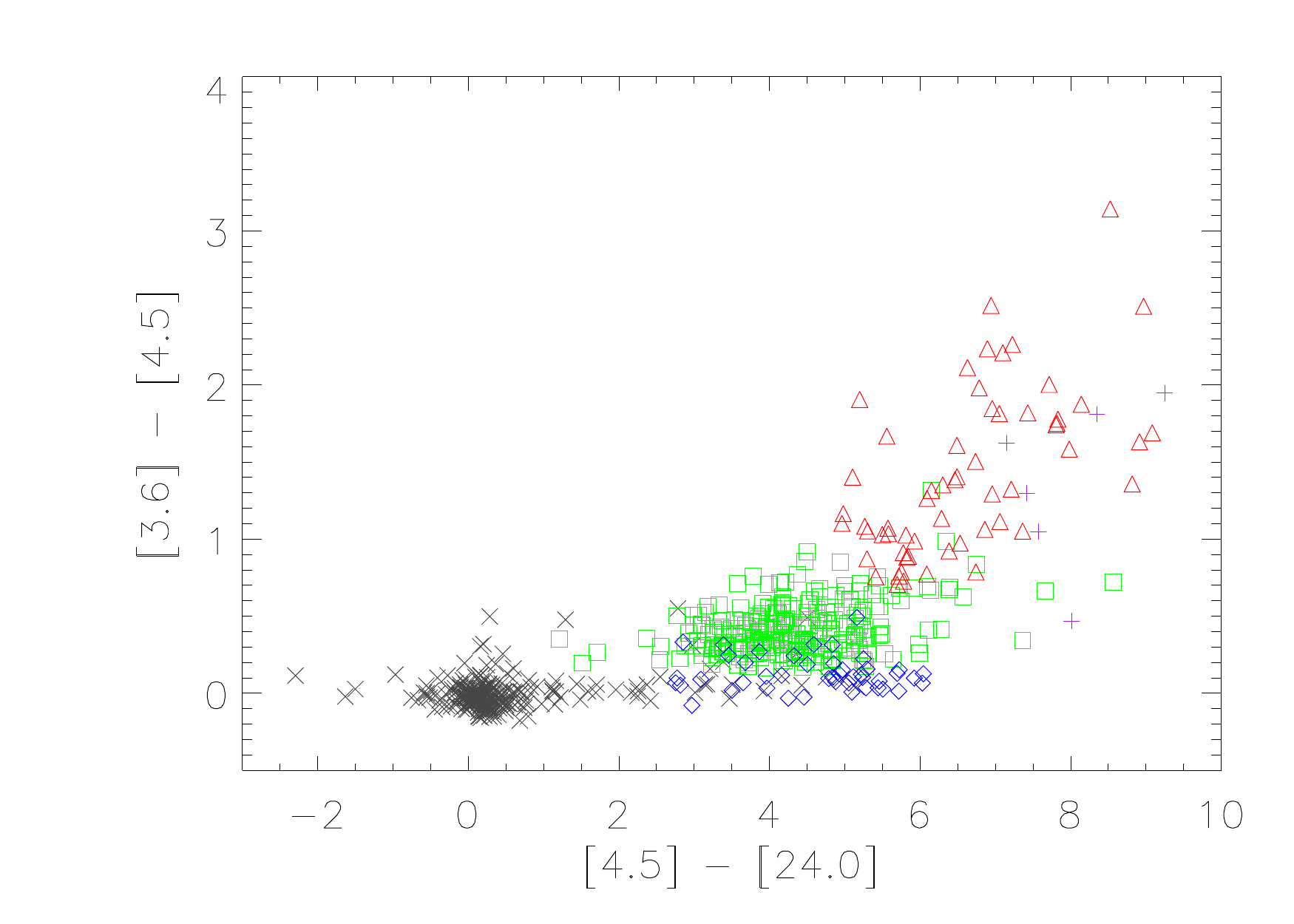} 
\caption{ \label{color_yso} Color-color diagrams showing
all classified YSOs/field stars. Red triangles are Class 0/I YSOs, green squares are Class II YSOs, blue
diamonds are transition disk objects, and gray crosses are Class III/field
stars.}
\end{figure}

Several classification schemes have been developed by different authors
in recent years to differentiate among protostars that are deeply embedded in gas and dust (Class 0), protostars
with dust emission predominantly from their envelope and disk (Class
I), pre-main sequence stars with
circumstellar accretion disks (Class II), stars with inner holes in
their accretion disks (Class II "transition disks"),
and diskless pre-main sequence (Class III) sources, as well as main
sequence stars in the line of sight. 
The bulk of emission from Class
0 sources is in the far-infrared/sub-mm, while Class I and Class II sources emit primarily
at mid-far infrared wavelengths. Both Class III YSOs and field stars exhibit
only photospheric emission, and thus cannot be distinguished using
only infrared photometry (see section 4). \citet{Allen2004} and \citet{Megeath2004} describe
a color-based YSO classification scheme using only IRAC data, and \citet{Muzerolle2004} describe a scheme using both IRAC and MIPS data. IRAC-only
methods of classification may miss very red sources or disks with inner holes while MIPS
24 $\mu$m images may not detect very faint YSOs, so a more involved
method is required to properly classify YSOs. \citet{Evans2009}
use extinction corrected IRAC and MIPS photometry and the slope of the mid-infrared spectral energy distribution (SED) to classify
YSOs in molecular clouds. \citet{Sung2008, Sung2009} and \citet{Hernandez2009,hernandez2008}, utilize X-ray and visible data to
distinguish between cluster and non-cluster members, then classify
YSOs based on their IRAC and MIPS photometry and the resulting shape
of the sources' SEDs. Since we only have 2MASS and \emph{Spitzer} data available for all of Mon OB1 East, we use the color-based classification scheme developed by \citet{Gutermuth2008}, amended by \citet{Gutermuth2009}, which uses a combination of infrared colors to distinguish between the different YSO classes. Our methods are summarized below, and the resulting classifications are shown in Figures \ref{image} and \ref{color_yso}.

\citet{Gutermuth2009} describe a three-phase method of classifying
YSOs that uses 7 band photometry from J-band to IRAC 8.0 $\mu$m to
classify each source and MIPS 24 $\mu$m photometry to confirm classifications and include additional YSOs where available, though data in all eight bands is not a requirement for classification. In this process, various methods are used to remove
contaminants such as active galactic nuclei (AGN), sources likely
dominated by emission from polycyclic aromatic hydrocarbons (PAH),
and shocked emission gas sources. Once these contaminants are removed,
the remaining sources are classified as YSOs or Class III/field stars.

In phase one, all point sources are analyzed, and potential YSOs  are identified as sources with data in all four IRAC bands with photometric uncertainties
$\sigma <$ 0.2 mag. Non-YSO contaminants which generally have IRAC
colors similar to those of YSOs are removed (see section 3.2) and
the remaining sources are classified using all four bands of IRAC
photometry. Sources with {[}4.5{]} \textendash{} {[}5.8{]} $>$ 0.7 mag, and {[}3.6{]}
\textendash{} {[}4.5{]} $>$ 0.7 mag are classified as Class 0/I protostars.
Sources with {[}4.5{]} \textendash{} {[}5.8{]} $>$ 0.5 mag, {[}3.6{]}\textendash{}{[}5.8{]} $>$ 0.35 mag
and {[}3.6{]}\textendash{}{[}4.5{]} $\leq$ 3.5({[}4.5{]}-{[}8.0{]}-0.5)+0.5 mag are
considered Class II YSOs, while the remaining sources are temporarily
classified as Class III/field stars.

In phase two, J, H, K, {[}3.6{]}, and {[}4.5{]} photometry are used
to identify additional YSOs from the presumed Class III/field star
population, as well as sources that lack detections at 5.8 or 8.0 $\mu$m, but have reliable 2MASS H and K photometry ($\sigma < 0.1$ mag) . The line-of-sight extinction to each source and color
excess ratios are determined using the reddening law developed by \citet{Flaherty2007} for all field stars, and dereddened K$_{S}$ \textendash{}{[}3.6{]}
and {[}3.6{]}\textendash{}{[}4.5{]} colors are determined using calculated color
excess ratios. If a source has infrared excess at 3.6 $\mu$m and 4.5 $\mu$m
after dereddening and accounting for photometric uncertainty, then
it is likely a Class 0/I or Class II YSO. Sources with no IR excess
at 3.6 $\mu$m and 4.5 $\mu$m are presumed Class III/field stars.

In phase three, 24 $\mu$m MIPS data are used to confirm prior classifications
of certain YSOs. Sources that were classified as Class III/field stars
in the previous phases but have excess emission at 24 $\mu$m ({[}5.8{]}
\textendash{} {[}24{]} $>$ 2.5 mag or {[}4.5{]} \textendash{} {[}24{]} $>$
2.5 mag) and {[}3.6{]} $<$ 14 mag are reclassified as transition disk objects.
Sources that lack detection in some or all IRAC bands, yet have very
bright 24$\mu$m photometry ({[}24{]} $<$ 7 mag and {[}IRAC{]} \textendash{}
{[}24{]} $>$ 4.5 mag, where {[}IRAC{]} is the photometric magnitude
for the longest wavelength IRAC detection obtained for that source)
are classified as Class 0/I YSOs. These sources are protostars that
are so deeply embedded in their natal material that reliable IRAC
photometry is unavailable in some bands. Finally, all
YSOs previously identified as Class 0/I protostars that have 24 $\mu$m
detections are reanalyzed to determine whether they are actually protostars
or highly reddened Class II YSOs. Sources with {[}5.8{]}\textendash{}{[}24{]}
$>$ 4 mag or {[}4.5{]} \textendash{} {[}24{]} $>$ 4 mag are considered bona fide Class 0/I protostars,
whereas sources that don't meet those requirements are reclassified
as highly reddened Class II YSOs. Figure \ref{color_yso} is a {[}3.6{]}\textendash{}{[}5.8{]}
vs. {[}4.5{]}\textendash{}{[}8.0{]} color-color diagram which displays the results
of applying the above classification criteria to all sources in our
data set.

\subsection{Isolating Contaminants}

Contaminants such as AGN, PAH emission sources, and {}``blobs''
of shocked emission exhibit infrared colors resembling those of YSOs.
The combination of near and mid-infrared data allows us to confidently identify
and remove these sources from the data set using the process described
by \citet{Gutermuth2009}. Diagnostic {[}4.5{]} \textendash{} {[}5.8{]}
versus {[}5.8{]} \textendash{} {[}8.0{]} and {[}3.6{]} \textendash{}
{[}5.8{]} versus {[}4.5{]} \textendash{} {[}8.0{]} color-color diagrams are used
to identify and remove PAH emission contaminants. Once PAH emission
sources are removed, we remove AGNs by exploiting the fact that even
though their MIR colors are very similar to those of YSOs, their 4.5 $\mu$m
magnitudes are much fainter ({[}4.5{]} $>$ 13.5 mag). A {[}4.5{]} versus
{[}4.5{]} \textendash{} {[}8.0{]} color-magnitude diagram is thus used to identify
likely AGN. We expect residual extragalactic contamination of $\sim$ 7 sources / degree $^{2}$.

\begin{figure}
\includegraphics[scale=0.5]{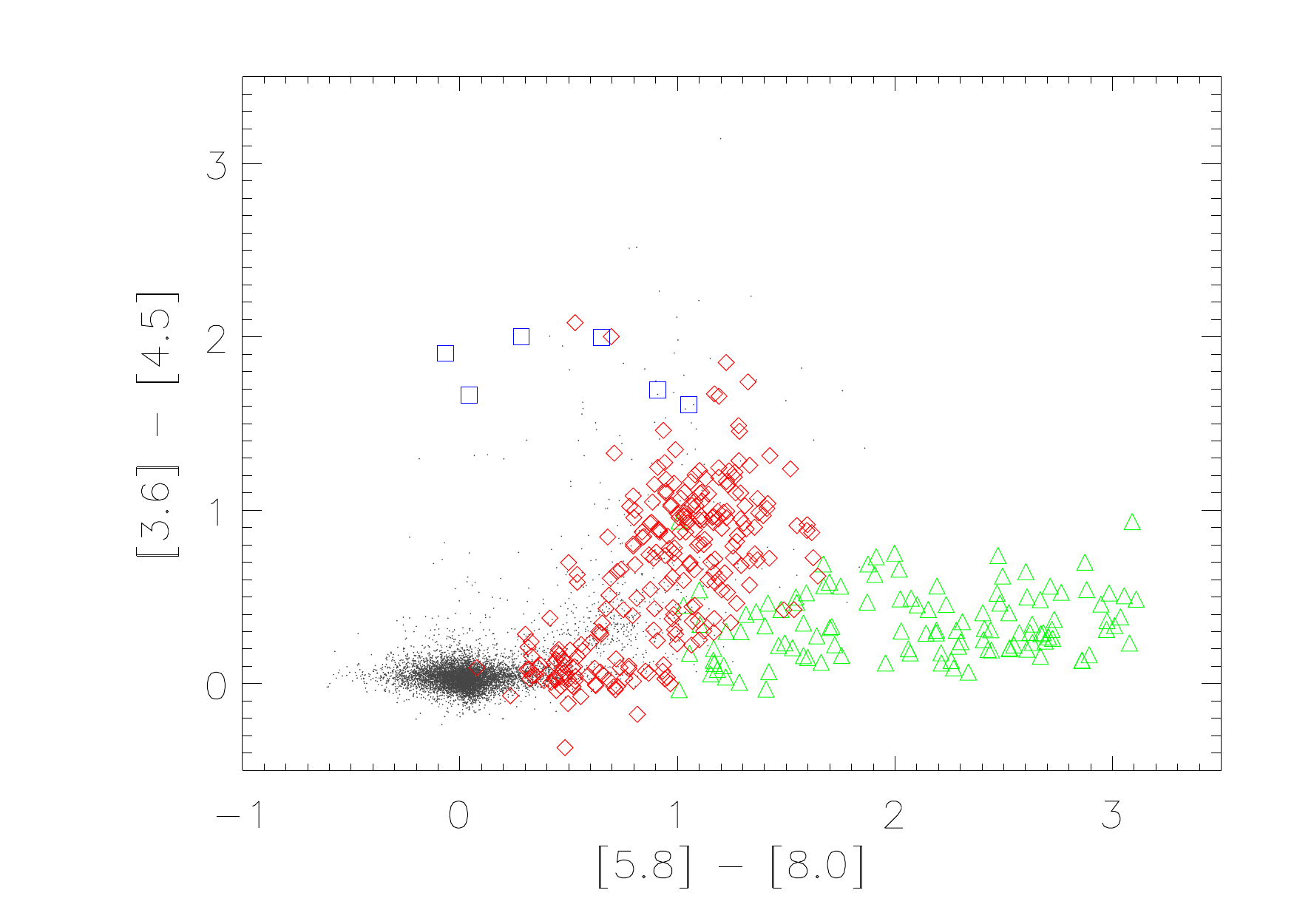}
\includegraphics[scale=0.5]{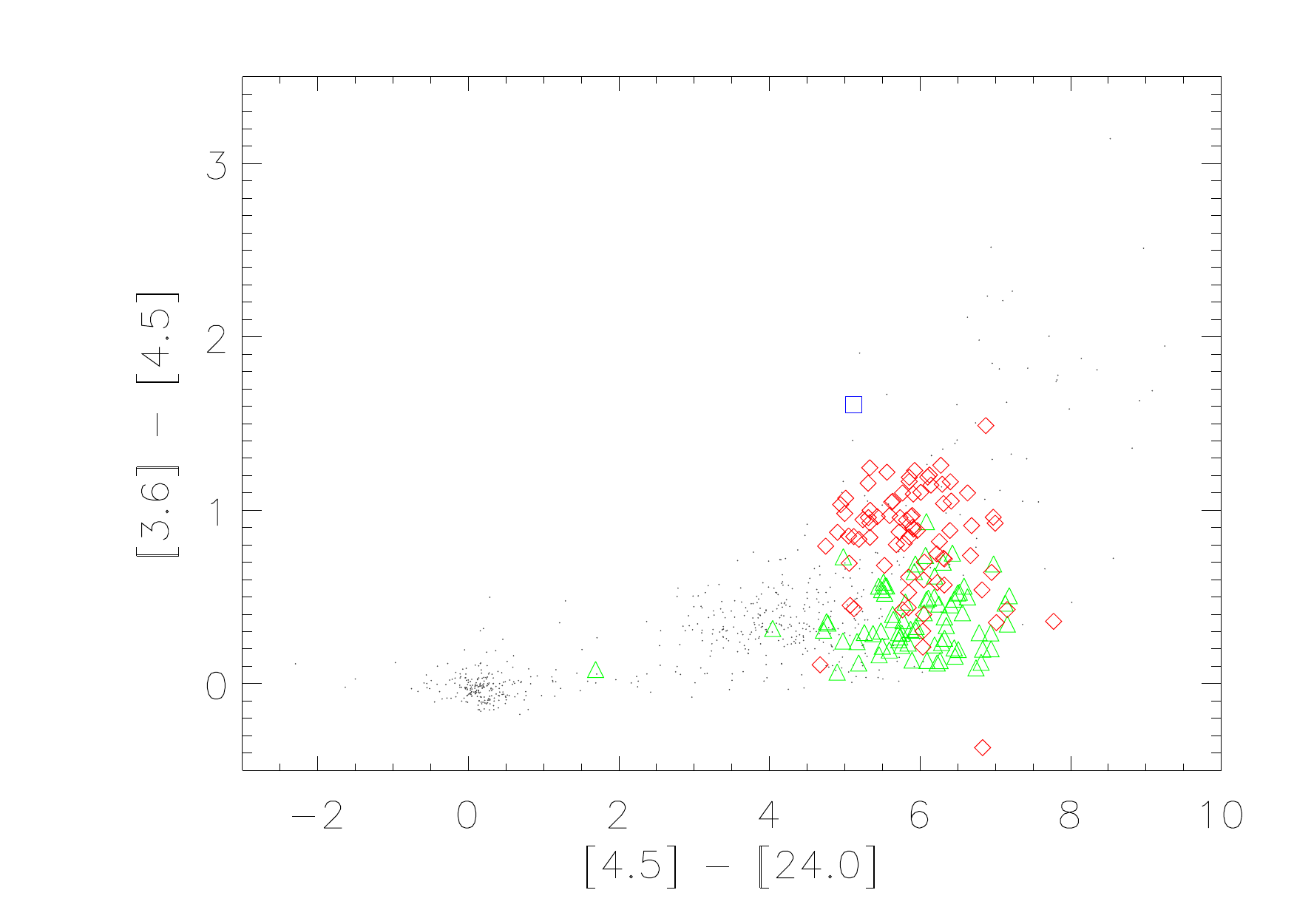}
\caption{ \label{color_contaminant} Color-color diagrams of
all previously classified sources (gray dots) with AGN (red diamonds),
PAH sources (green triangles), and blobs of shocked emission gas (blue
squares) overplotted. The contaminants have similar colors to those
of YSOs but are distinguished based on the procedure described in
the text. All sources identified as contaminants are removed from
the data set.}
\end{figure}

We remove the final contaminant, unresolved blobs of shocked emission
gas tracing high velocity outflows that interact with the cool molecular
cloud, by exploiting the fact that these sources have large 4.5 $\mu$m
excess due to shocked molecular hydrogen line emission in this particular
band \citep{Smith2006}. Using {[}3.6{]} \textendash{} {[}4.5{]}
and {[}4.5{]} \textendash{} {[}5.8{]} colors, we can thus determine
whether a source is likely dominated by shock emission and remove
it from our data set. Figure \ref{color_contaminant}  shows a color-color diagram of all classified contaminants plotted
over all previously classified YSOs.

\subsection{Counts of Identified YSOs and Contaminants}

After running custom IDL programs 
to classify the YSOs based on the above recipes  \citep{Gutermuth2009}, we identify 10,454
point sources as potential YSOs out of the original 151,350 sources in the line of sight to Mon OB1 East. Of these 10,454 potential YSOs, 7,430 were detected in all four IRAC bands. After removing 117 likely PAH emission sources,
250 likely AGN and six likely blobs of shocked emission gas from the total sample, 10,081
objects remain. Of these 10,081 objects, 105 are classified as Class 0/I, 434 are classified as Class II, and 45
are classified as transition disk objects, for a total of 584 disk sources.  9,497 are classified
as Class III/field stars, and a statistical estimate of the likely number of Class
III sources (as opposed to field stars) is discussed in the following
section. Of the 3,024 sources that lack data all IRAC bands, 13 are still able to be classified as class O/I and 18 as class II sources, and the rest have enough reliable photometry to be classified as YSOs with disks or Class III/field stars. We present photometric data and classifications for the 10,454 YSOs
and contaminants in Table 1. Each source is identified by a \emph{Spitzer}
source detection number and position in columns 1-3. Magnitudes and
uncertainties in the four IRAC bands and 24 $\mu$m MIPS are listed
in columns 4-8 and our classification for each source is listed in
column 9. Classification from \citet{Sung2008,Sung2009} is listed in column 10 for all sources in our data set that spatially coincide with sources presented in their work. Figure \ref{image} shows the location of all classified disk sources towards Mon OB1 East.

\subsection{Estimating the Number of Class III Sources}

To estimate disk fractions for different regions of Mon OB1 East,
a distinction must be made between diskless Class III YSOs and field
stars in the line of sight. IRAC and MIPS data are available for a
majority of the cloud, but since both Class III objects and field
stars display pure photospheric emission, such a distinction cannot
be made using only \emph{Spitzer} colors. Instead, we use a K$_{S}$-band
number count method to estimate the number of field stars in a given
direction towards the cloud, and thus a disk fraction for the entire cloud.

\begin{figure}
\includegraphics[scale=1]{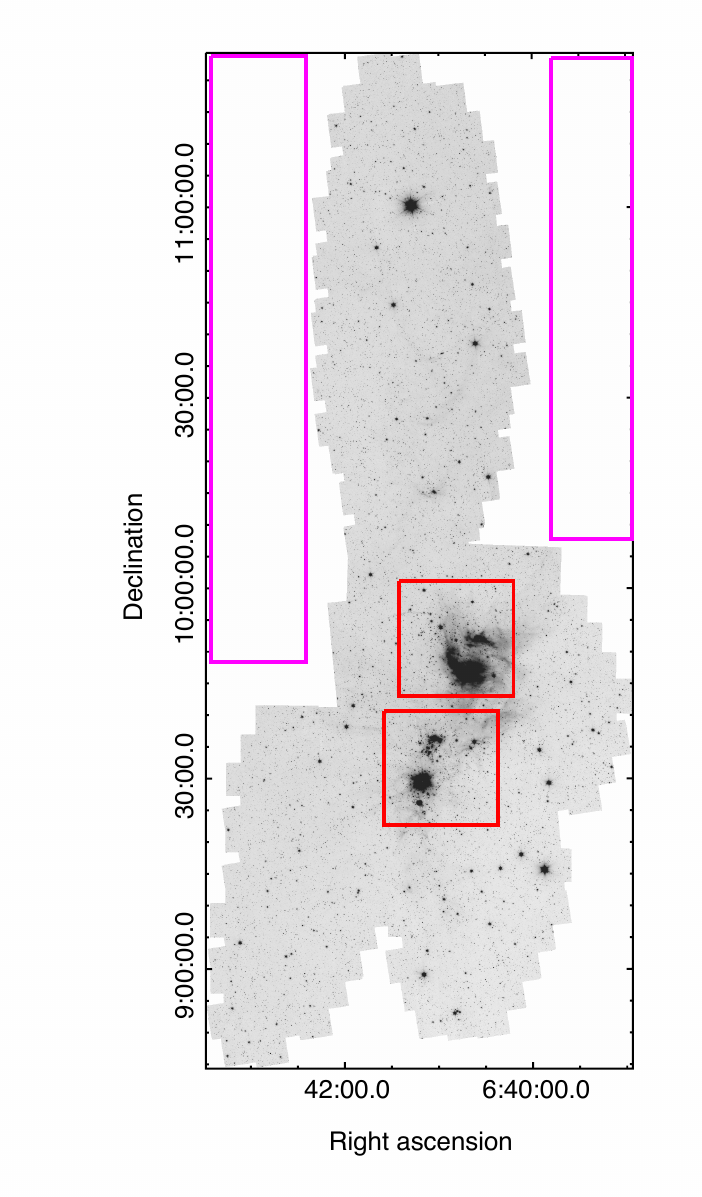}
\includegraphics[scale=0.98]{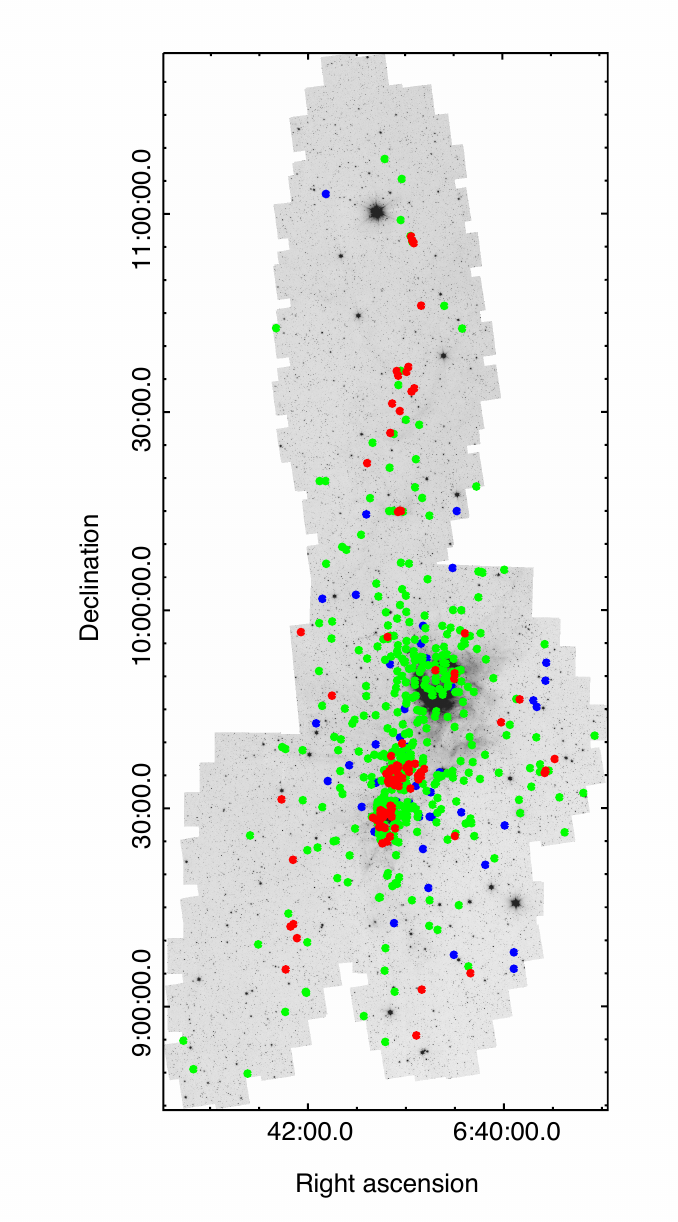}
\caption{ \label{image}  Left: \emph{Spitzer} 4.5 $\mu$m inverted grayscale image of Mon OB1 East. NGC 2264 is outlined by the red squares which encase the S Mon region (top square) and Cone region (bottom square) of \citet{Sung2008}. All Spitzer data outside of the two red boxes represent Mon OB1 cloud. The position of the two reference regions used to estimate the number of field stars towards Mon OB1 East are outlined by the magenta boxes to the east and west of the cloud. We used 2MASS data for all sources within the magenta boxes to estimate the number of field stars. Right: Same \emph{Spitzer} 4.5 $\mu$m inverted grayscale image of Mon OB1 East but with Class 0/I (red), Class II (green) and transition disk (blue) sources overlaid.}
\end{figure}

To estimate the number of Class III sources within Mon OB1 East,
we consider field star densities in neighboring regions to determine
the likely number of field stars within Mon OB1 East as a function of magnitude. We obtain 2MASS data for all point sources located in two ``reference" regions to the East and west of the northern portion of Mon OB1 East (Figure \ref{image}) to estimate field star densities. The reference fields are located in close proximity
to Mon OB1 East and, therefore, the background surface density of stars
toward the cloud is assumed to be the same as that of the
reference fields. Since extinction varies between NGC 2264 and the rest of Mon OB1 East, we estimate separately the number of field stars towards NGC 2264, and the rest of the cloud (hereafter Mon OB1 cloud). We consider the NGC 2264 region as two squares $\sim$ 0.3 x 0.3$^{\circ}$ (Figure \ref{image}) that encompass all three sub clusters, and roughly coincide with the S Mon and Cone regions of \citet{Sung2008} for ease of comparing our results to those of \citet{Sung2009}. The Mon OB1 cloud is considered to be the region covered by the \emph{Spitzer} IRAC data, minus the two NGC 2264 regions. 

To estimate the number of field stars towards Mon OB1 East,
we use the number counts of all sources (classified YSOs and Class III/field stars)
in each region (NGC 2264 and Mon OB1 cloud), along with those in the two reference regions, that have 2MASS K$_{S}$ magnitudes less than 13.5 mag and K$_{S}$
uncertainties less than 0.2 mag, into K$_{S}$-band magnitude histograms
(KMH) as described by \citet{Gutermuth2008}. K$_{S}$-band magnitudes
are available for 84\% of the \emph{Spitzer} classified sources with
disks, and 99\% of the Class III/field sources towards Mon OB1 East. 

Disk sources exhibit a larger flux in the infrared than pure photospheric sources, and thus
can be detected to fainter magnitudes and higher extinctions. To account for this bias,
we follow the method of \citet{Allen2012} and apply an extinction limit and dereddened J-band magnitude limit to ensure the detection of sources both with and without IR-excesses in the combined 2MASS and Spitzer data.   The J-band was chosen since it is the IR-band least affected by IR-excesses. We use in our analysis only sources whose 
position on a J vs. J-H color-magnitude diagram indicate an extinction of A$_{V} <$  6 and are
brighter than a de-reddened J-band magnitude of 14.6. This magnitude and reddening vector corresponds 
to a 0.2 M$_{\odot}$ star at a distance of 760 pc using the models of \citet{Baraffe1998}.  \citet{Allen2012} determine these values to be appropriate magnitude extinction cutoffs for the Cep OB3b cluster which was is at a similar distance and is observed with a similar sensitivity After applying this cut to the sources towards Mon OB1 East , we are left with a total of 341 disk sources (Class 0/I, II, or transition disk), and 5952 Class III/field stars. 

\begin{figure}
\includegraphics[scale=1]{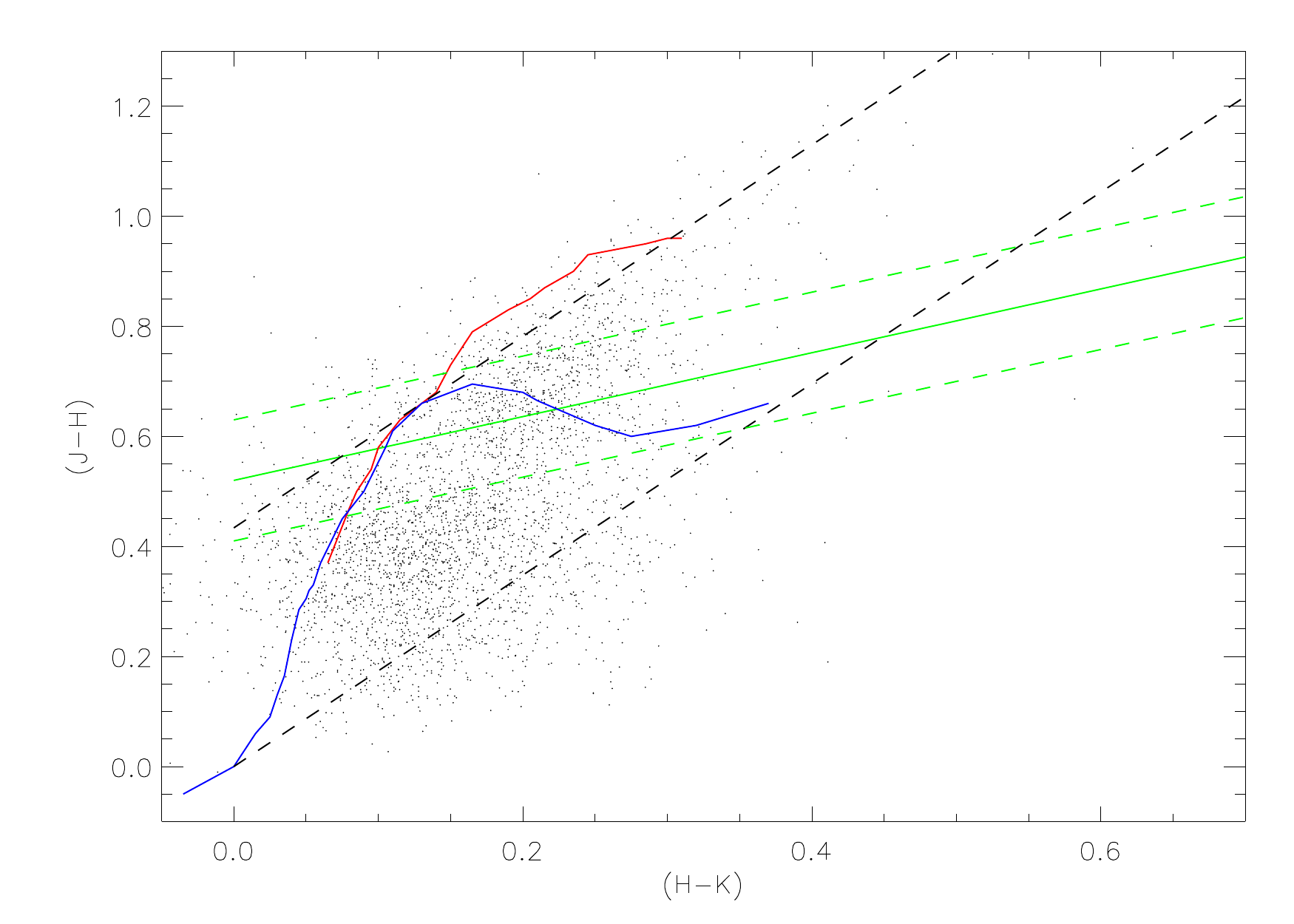}
\caption{\label{2mass_yso} [J-H] vs [H-K] diagram of stars in the two field regions combined from 2MASS data.
The data are compared with the colors of giants
(red locus) and dwarfs (blue locus) from \citet{BessellBrett1988} and
the classical T-Tauri star locus (green locus) from \citet{Meyer1997}. Dashed green lines represent 1$\sigma$ errors
in the T-Tauri locus. Dashed black lines indicate reddening vectors from \citet{Cohen1981}
for giants (top) and M stars (bottom).}
\end{figure}

It is possible that Mon OB1 East extends beyond the region covered by \emph{Spitzer}, and thus our reference
regions could contain sources with disks. We use 2MASS data to perform a simple check for disk sources in the reference field. We construct a {[}J-H{]} vs. {[}H-K$_{S}${]}
color-color diagram (Figure \ref{2mass_yso}) of all sources in the reference field used in the disk fraction calculation and
compare their NIR colors to those of classical T-Tauri stars \citep[green locus;][]{Meyer1997}, giants (red locus) and dwarfs 
\citep[blue locus;][]{BessellBrett1988}. Most sources appear to follow the same trend
as dwarfs or giants, and only about 5\% fall within the boundaries of the classical
T Tauri star (CTTS) locus. These sources could be YSOs, or more likely reddened field stars. To further examine the reference regions for contamination, we calculated the average stellar surface density of all 2MASS sources in the reference field to check  for regions of clustered, evolved YSOs that belong to Mon OB1 East. The average was 25 sources/ pc$^{2}$, with a maximum density of $\sim$ 70 sources/ pc$^{2}$ nearest Mon OB1 East. We also calculated these values for the reference regions used by \citet{Teixeira2012}, who conducted a similar study of NGC 2264 (see section 4), that were $\sim1^{\circ}$ square and obtained almost identical results. This suggests that our reference regions are not significantly contaminated by members of Mon OB1 East. Thus, we conclude that our reference regions are a good representation of the field star background towards the cluster. 

\begin{figure}
\includegraphics[scale=1]{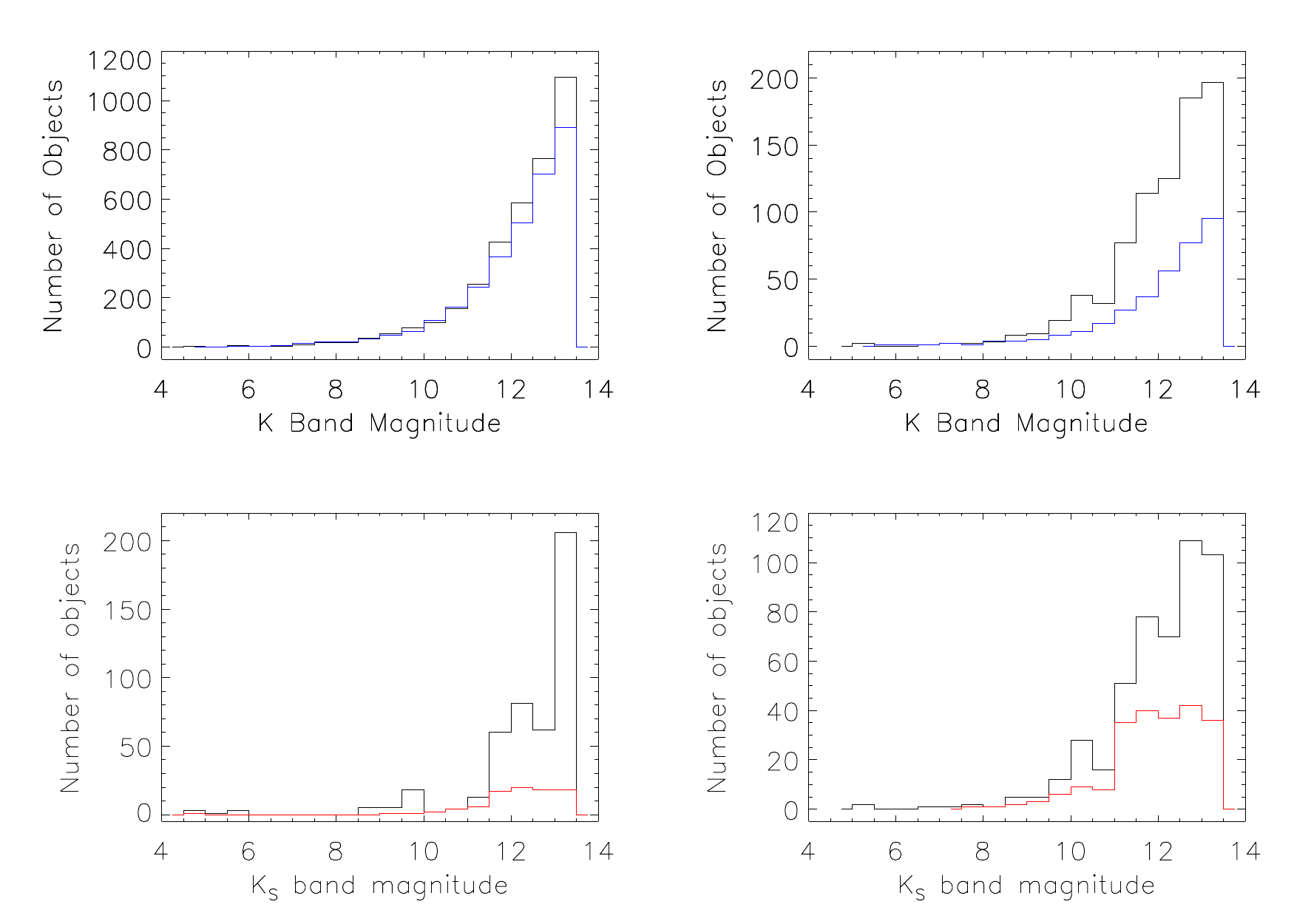}
\caption{\label{k_histogram} K$_{s}$-band magnitude histograms for Mon OB1cloud (left), and NGC 2264 (right).
 Top row: Black histogram represents all sources towards the region and the blue histogram represents the number of sources in the 
reference fields. Bottom row: Black histogram represents the total YSO count (sources with disks plus
Class III) after field stars have been statistically removed. Red
histograms represent counts of \emph{Spitzer} classified disk sources.}
\end{figure}

Now that we have checked for potential YSOs in the reference region, we can estimate the 
total number of Class III sources within Mon OB1 East . 
We calculate the  K$_{S}$-band extinction of all point sources for each region and reference field using 2MASS data. To account for differing extinctions, we calculate the mean extinction of the two reference regions from the extinction towards each individual source, then de-redden the eastern reference
region by the difference in average extinctions. This results in the two reference fields having approximately the same average extinction. Sources in this region are then artificially extinguished by the mean extinction
value for NGC 2264 and Mon OB1 cloud regions individually. After
correcting for mean extinctions, we scale the area, and thus number of sources, to that of each region. We then estimate the total number of YSOs (both disk sources and Class III) in NGC 2264 and Mon OB1 cloud by subtracting the KMH of the scaled reference
field from that of all the stars in each region. The resulting KMHs are shown in Figure \ref{k_histogram}. We find 347 field stars towards NGC 2264, and 3203 field stars towards Mon OB1 cloud. 

\section{Disk Fraction}

To calculate a disk fraction, we sum the number of sources in each
bin of the KMH of classified infrared excess sources (all Class O/I and Class II YSOs) and divide by
sum of the total number of sources in each bin of the background subtracted KMH of all sources (all YSOs + estimated number of Class III sources).  Table 2 gives the number of \emph{Spitzer} identified excess sources that satisfy our dereddened J-band magnitude, extinction and K$_{S}$-band magnitude criteria, the calculated number of Class III sources,
and the resulting disk fraction. We find disk fractions 
of 45 $\pm$ 4\% for NGC 2264 and 19 $\pm$ 5\% for Mon OB1 cloud. Appendix A describes how we calculated the uncertainty associated with the disk fractions. 

\subsection{Comparison To Previous Work}

A series of papers by \citet{Sung1997,Sung2004, Sung2007, Sung2008, Sung2009} cover in detail the selection criteria
they used to determine cluster membership within NGC 2264. Cluster members were determined
by \citet{Sung2008, Sung2004} using visible and near-infrared color-color diagrams,
H-$\alpha$ equivalent widths, and X-ray photometry of the sources.
\citet{Sung2009} used point spread function (PSF) photometry to analyze the \emph{Spitzer}
data, and a combination of SED fitting
and color-color diagrams to classify cluster members as YSOs with
disks. They identify 21,991 \emph{Spitzer} sources in a roughly 40' x 80' region of Mon OB1 East , and from these data classify YSOs with circumstellar disks/envelopes (Class 0/I and 
Class II sources) in order to understand the spatial density and
distribution of YSOs within their natal material. Using H-$\alpha$
and X-ray data, they calculate the number of diskless pre-main sequence stars (Class III) within the cluster 
and thus the fraction of all YSOs with disks for the regions around S Mon and the Cone nebula. They find a disk 
fraction of 38\% for a region roughly coincident with our NGC 2264 region, a value less than our estimated disk fraction of 45 $\pm$ 4\%. 

This discrepancy may arise from differing YSO classification schemes, and stricter classification criteria. \citet{Sung2009} identify 122 disk sources in the NGC 2264 region, whereas we identify 220.  They require the sources to be found in a pre-ms locus defined in a I vs R-I diagram and then use a combination of mid-IR colors  and SED slopes to identify sources with disks. Our classification is based solely on the Spitzer data, which may be why we find roughly twice as many disk sources in the same region. The lower disk fraction reported by \citet{Sung2009} may come from the requirement that the sources are on the pre-ms locus established in the I and R-band photometry, and therefore they do not include more embedded sources. Since we do not have equivalent visible light photometry for all of Mon OB1 East, we cannot determine cluster membership using a similar techniques to \citet{Sung2004, Sung2008, Sung2009},and we must rely on our statistical determination of disk fraction.

\citet{Teixeira2012} also estimate the number of YSOs in a region slightly larger than NGC 2264, as defined in this work.
They use \emph{Spitzer} data to calculate the slope of the SED for each object, and thus determine if the object is a Class I YSO,
flat spectrum source, source with a thick disk, or source with an anemic disk. After correcting for extinction,
they use FLAMINGOS data and a K$_{s}$-band magnitude histogram to estimate the total number of sources which belong to the cluster, a method very similar to ours. They find a total of 1436 $\pm$ 242 YSOs in the cluster, 372 of which show \emph{Spitzer} infrared-excess, resulting in a disk fraction of approximately 26\%. This value is considerably smaller than our determined disk fraction around NGC 2264, but \citet{Teixeira2012} have deeper infrared observations.  They have determined the total number of sources from an analysis of K-band luminosity function, and thus this count may include many sources that do not have the required IRAC photometry needed to identify disks.  Thus, 26\% should be considered a lower limit to the disk fraction determined from their data.  Furthermore, they calculate the disk fraction for a larger portion of the cloud than just our designated NGC 2264 region. Based on our analysis, it is clear that beyond the sub-clusters within NGC 2264, the number density of disk sources drops dramatically (see section 6). Therefore, \citet{Teixeira2012} may be including a large number of Class III sources from an area with a low number of YSOs, thus resulting in a small disk fraction. 

\section{Extinction Map of the Mon OB1 East Molecular Cloud}

We derive an A$_{K_{S}}$ extinction map by dereddening (H\textendash{}K$_{S}$)
colors of 151,350 point sources toward the Mon OB1 East using a method
discussed by \citet{Gutermuth2005}. To summarize, the line of sight
extinction to each point on a 3.5\arcsec  grid throughout
the region surrounding the central point is estimated using a variation
of the nearest neighbor technique. For each point, the average and
standard deviation of the (H\textendash{}K$_{S}$) colors of the 20 nearest stars
to each grid point are calculated and any outliers with an (H\textendash{}K$_{S}$)
color greater than 3$\sigma$ from the mean are rejected until the
mean converges. We adopt the NIR reddening law of \citet{Flaherty2007} for the seven bands 
(J-8.0 $\mu$m) and convert the (H\textendash{}K$_{S}$)
value at each point to A$_{K_{S}}$ using A$_{K_{S}}$ = 1.82{[}(H-K$_{S}$)$_{obs}$\textendash{}
(H\textendash{}K$_{S}$)$_{int}${]}, where (H\textendash{}K$_{S}$)$_{int}$ = 0.2 is adopted
as the average intrinsic (H-K$_{S}$) color for field stars. We find an average extinction of the molecular cloud coincident with the NGC 2264 region to be A$_{K_{s}} \sim$ 0.45, and A$_{K} \sim$ 0.22 mag for Mon OB1 cloud. \citet{Teixeira2012} calculate the extinction towards NGC 2264 and, converting their measurements of A$_{V}$ to A$_{K_{s}}$ using the interstellar extinction law from \citet{Cardelli1989}, they find a mean of A$_{K_{s}} $ = 0.696 and median of A$_{K_{s}}$ = 0.507 mag. While their extinction measurements are slightly larger than ours, if we use their average value in our calculations of disk fraction for NGC 2264, we obtain a disk fraction of 42\% which is comparable to our derived value of 45 $\pm$4\%.

\begin{figure}
\includegraphics[scale=0.6]{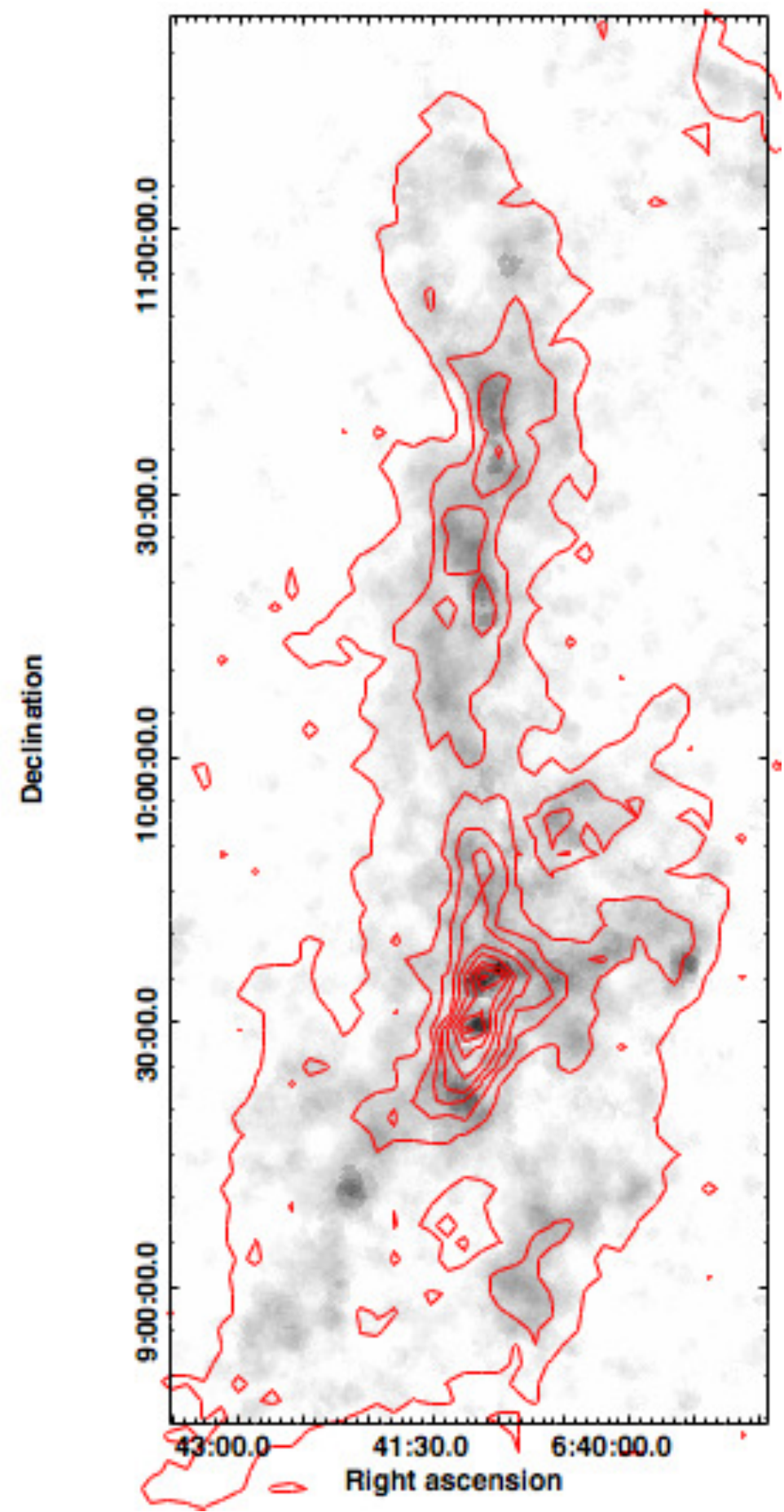}
\caption{\label{CO} Grayscale A$_{K_{S}}$ extinction map of Mon OB1 East with $^{13}$CO contours from
\citet{Reipurth2004II} overlaid. Grayscale extends from A$_{K_{S}} = 0.2$
mag (white) to 2.0 mag (black). The $^{13}$CO emission has a maximum
of 153.93 K Km s$^{-1}$and contours ranging from 10\% to 90\% of
the maximum in increments of 8\%.}
\end{figure}

We compare our resulting extinction map with $^{13}$CO map contours from \citet{Reipurth2004II} in Figure \ref{CO}.
 The A$_{K_{S}}$  extinction map shows
a distribution similar to that of the $^{13}$CO map, especially for NGC 2264. Any slight differences
between the two maps may be due to variations in
gas opacity, freezeout of $^{13}$CO, the lower spatial resolution of the $^{13}$CO map, or variations in excitation temperature \citep{Ripple2013}. The A$_{K_{S}}$ extinction map provides greater spatial detail, especially
in densely populated areas.

\section{Spatial Distributions of YSOs and Molecular Gas}

\begin{figure}
\includegraphics[scale=1.5]{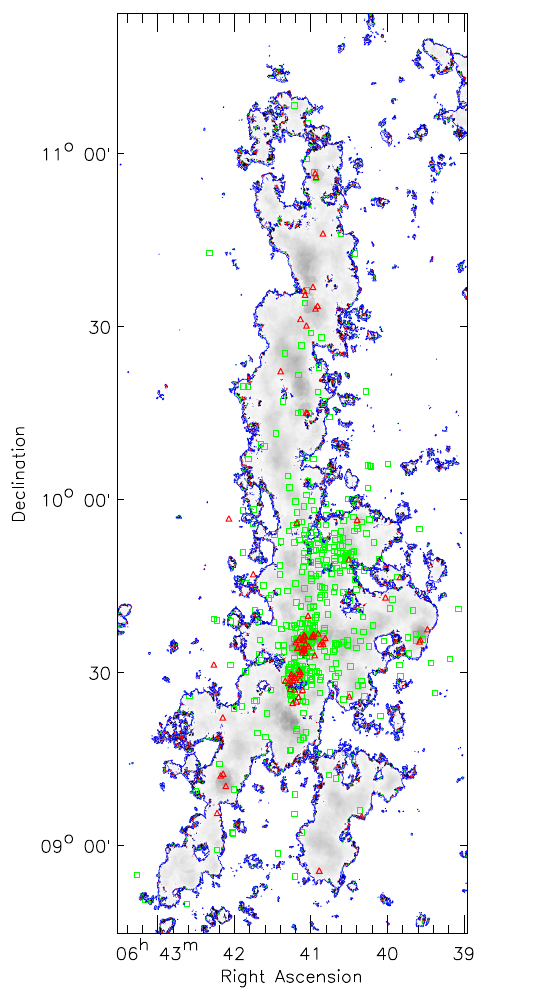}
\caption{\label{extinction} Same extinction map as figure 7 with \emph{Spitzer} identified disk
sources overplotted. Red triangles are Class 0/I sources, green squares
are Class II sources, and blue contour represents A$_V$=1.}
\end{figure}

Assuming that the gas to dust ratio is a constant, extinction-derived
dust column density is a proxy for molecular gas column density. Our
extinction map can then be interpreted in terms of the gas mass column
density, assuming $\frac{A_{V}}{N_{H_{2}}}=10^{-21}$mag-cm$^{2}$ \citep{Cox2000}, by multiplying it by 2.8. This includes both the mass of hydrogen and the assumption that He/H is $\sim$10\% by number. Figure
\ref{extinction} shows the same grayscale A$_{K_{S}}$ extinction map of NGC 2264
as in Figure \ref{CO}, but now all Class 0/I (red triangles) and Class II (green
squares) YSOs are overlaid. YSO number density and extinction both
peak within NGC 2264, where active
star formation is likely occurring. 

\subsection{Examining the Correlation Between Gas Column Density and YSO Surface
Density}

Many have shown \citep[e.g.][]{Carpenter2000, Allen2007} that star
forming regions in molecular clouds tend to contain both dense clustered
regions of star formation surrounded by a more diffuse population
of YSOs. The youngest YSOs trace the dense gas of their natal
cloud \citep [e.g.][] {Gutermuth2005}.  A number of authors \citep[e.g.][Megeath et al. 2014, submitted.]{Hillenbrand1998} 
show that diffuse distributions of YSOs, like those surrounding
NGC 2264, join smoothly with the dense clusters and follow the filamentary morphology of their parental clouds. Simulations suggest that this type of sub-clustering may be formed via turbulent fragmentation \citep{Wang2010, Bate2012}

\begin{figure}
\includegraphics[scale=1]{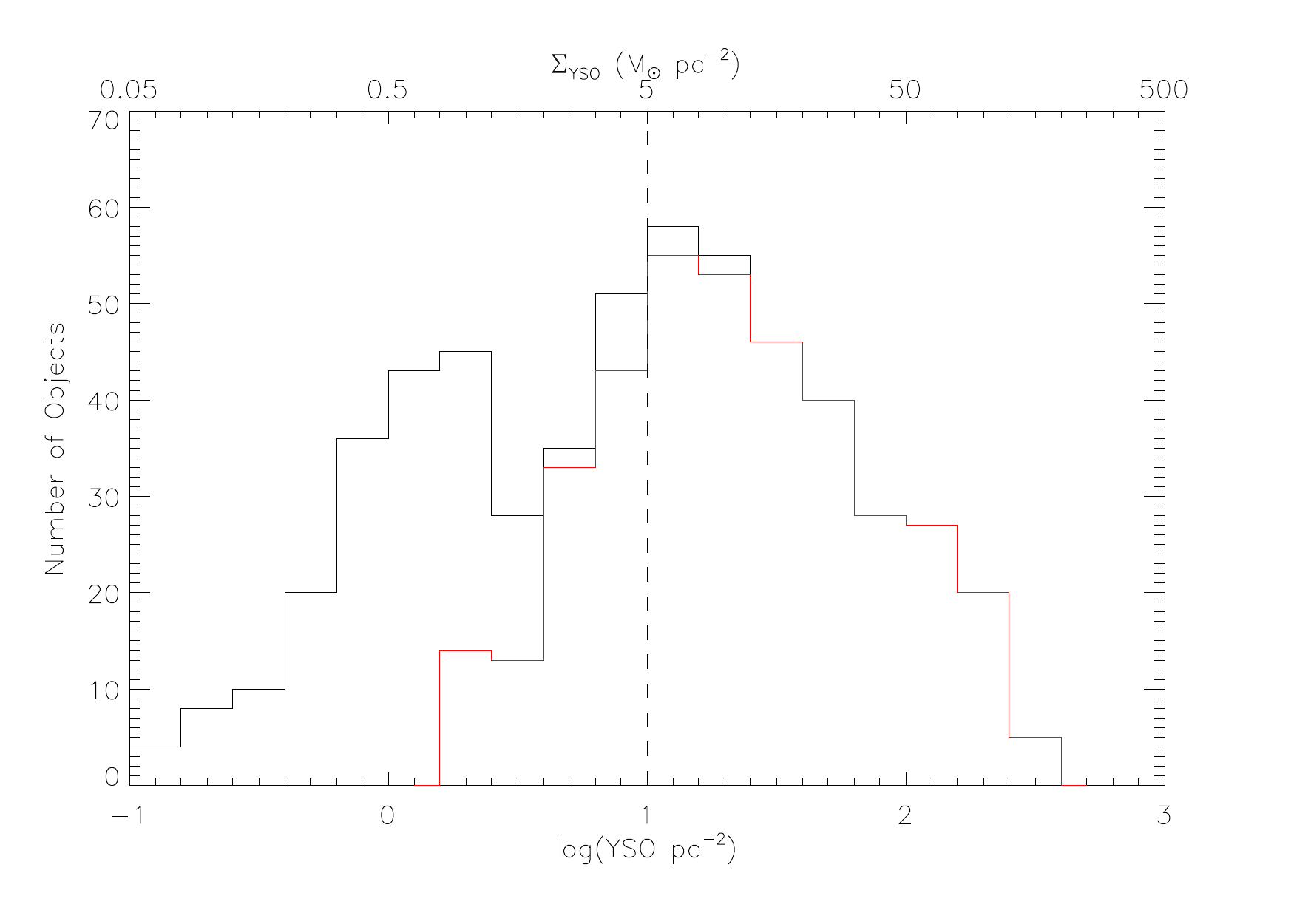}
\caption{\label{number} Histogram of the number of YSOs per square parsec for Mon OB1 East (black) and just NGC 2264 (red). The vertical dashed line at a density of 10 pc$^{-2}$ represents the threshold for clouds containing embedded populations (above this density), and clouds with a more dispersed population (below this density; Megeath et al. 2014, submitted.)   } 
\end{figure}

We can study the cloud demographics by observing the YSO spatial density throughout Mon OB1 East, to determine how clustered versus dispersed the cloud is (Megeath et al. 2014, submitted.). Figure \ref{number} shows the log(YSO/pc$^{2}$) for Mon OB1 East (black), and just NGC 2264 (red). YSO density was calculated using the distance to the 11th nearest neighbor. Although is no clear density threshold between clustered and lower density dispersed regions, Megeath et al. 2014 (in prep.) find that a 10 pc$^{-2}$ density threshold approximately distinguishes dense clusters in Orion with the more distributed regions in Taurus. We find that the nearest neighbor densities in NGC 2264 cluster peaks above this threshold and most of the YSOs in NGC 2264 are found at densities above 10 pc$^{-2}$.  In contrast, the reminder of the cloud the YSOs are found primarily in regions with densities below 10 pc$^{-2}$; consistent with the NGC 2264 region being dominated by clustered star formation while the star formation in the remained of the cloud is more spatially distributed.

The distribution of YSOs as compared to molecular gas density can also reveal concentrated regions of star formation within a GMC.
\citet{Gutermuth2011} examined 8 entire molecular clouds and found
a correlation between the local gas mass density and the YSO mass
surface density. The molecular gas mass column density ($\Sigma_{gas}$) and YSO mass
surface density ($\Sigma_{*}$) throughout these clouds appear to follow a power law of the form $\Sigma_{*}\propto\Sigma_{gas}^{N}$, where N ranges from 1.87 to 2.67. 
This implies that the range of YSO surface densities apparent in Figure \ref{number} are due to the range of gas column densities within Mon OB1 East.  They further imply that the star formation efficiency varies with $\Sigma_{gas}$, with low density regions having star forming efficiencies of a few percent while dense, clustered regions have star formation efficiencies of 25\%.  The physical explanation for this relationship is not clear. Assuming continuous star formation, this relationship would imply that the star formation rate scales with $\Sigma_{gas}^{N}$. A power law with N$\sim$ 2 may be explained by a simple thermal fragmentation model of isothermal, self-gravitating layers of dense gas if the mass and duration of fragmentation and collapse is independent of density  \citep{Gutermuth2011}. Thus, the higher YSO densities in regions of higher molecular gas mass column density may due to thermal fragmentation.  Alternatively, a star formation rate proportional to the free fall time  could give an equivalent relationship if the star formation efficiency were constant \citep{Parmentier2013}.

\begin{figure}
\includegraphics[scale=1]{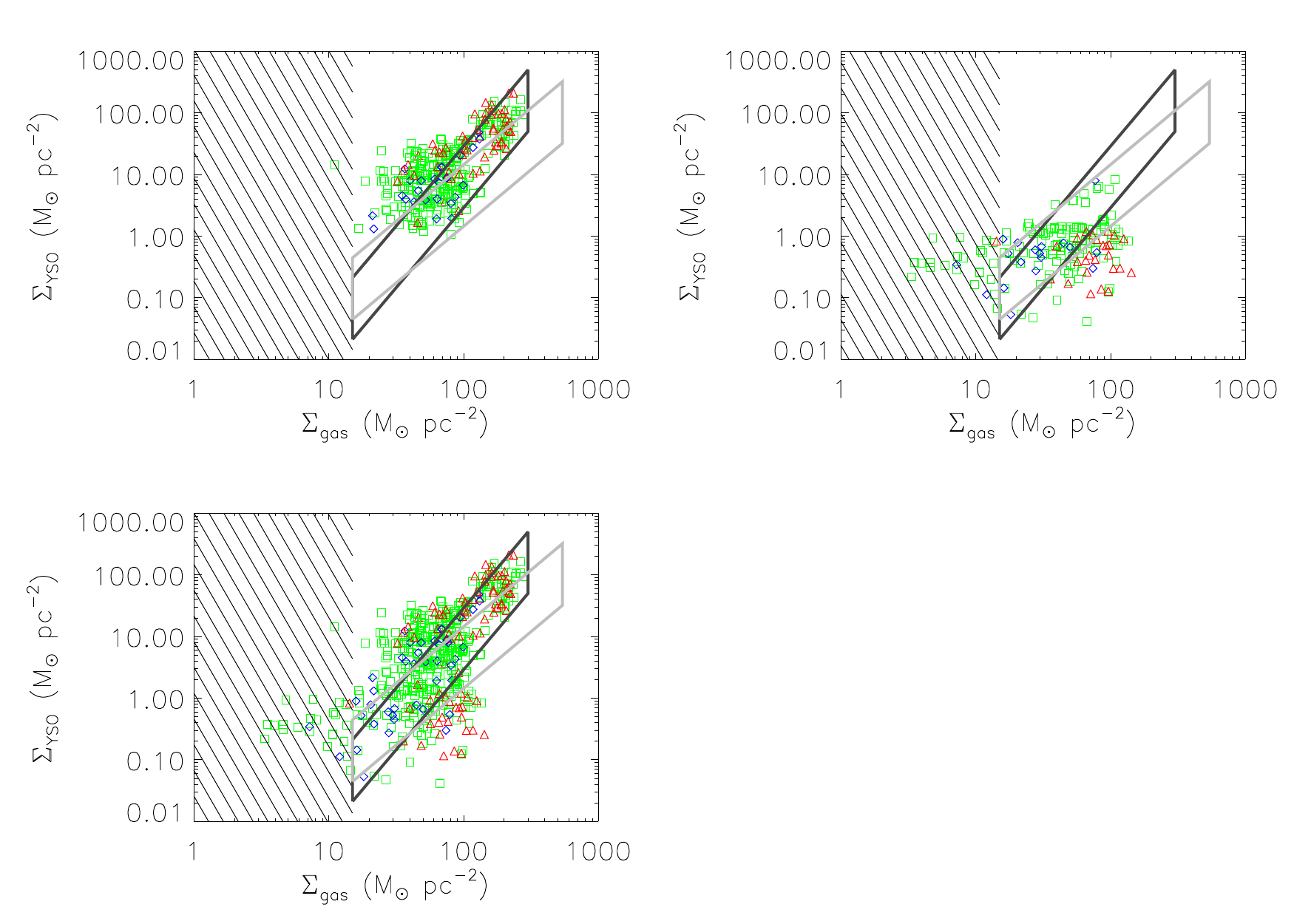}
\caption{\label{surface_density} YSO mass surface density vs. molecular gas mass column
density of all \emph{Spitzer} identified Class 0/I (red triangles), Class
II (green squares) and transition disk sources (blue diamonds) for NGC 2264 (top left), Mon OB1 cloud (top right), and all of Mon OB1 East (bottom). Overplotted
are polygons marking the fit to, and estimated spread in, data of
the Mon R2 molecular cloud (black; N=2.67 $\pm$ 0.02), and Ophiuchus molecular cloud (gray; N=1.87 $\pm$ 0.03)
from \citet{Gutermuth2011}. Hatched regions show where AV < 1 and the gas column density cannot be reliably determined from the near-IR extinction.}
\end{figure}

To assess whether Mon OB1 East follows a similar scenario to other GMCs,
we plot $\Sigma_{*}$ vs $\Sigma_{gas}$ for all Class 0/I, II and transition
disk sources in NGC 2264, Mon OB1 cloud, and all of Mon OB1 East (Figure \ref{surface_density}). To determine the gas and YSO surface density, we follow the same procedure as \citet{Gutermuth2011}. A YSO centered 11th nearest neighbor surface density analysis was conducted on \emph{Spitzer} classified YSOs to determine the YSO surface densities, as was done for Figure \ref{number}. The hatched region in Figure \ref{surface_density}
marks the area on the diagram where A$_{V} <$ 1, i.e. consistent
with A$_{V}$ = 0, given uncertainties.

In Figure \ref{surface_density} we overplot two polygons representing the fit to, and
estimated spread in, the $\Sigma_{YSO}$ vs. $\Sigma_{gas}$ relation on the Mon R2 molecular cloud (black; N=2.67 $\pm$ 0.02), and
Ophiuchus molecular cloud (gray; N=1.87 $\pm$ 0.03) from \citet{Gutermuth2011}.  Mon OB1 East follows a trend similar to Mon R2 (slope 2.67$\pm$ 0.02) but exhibits considerable spread in the correlation (Pearson index of 0.74),
similar to that observed for Orion \citep{Gutermuth2011}. Mon OB1 cloud alone does not clearly follow the trend of Mon R2 or Ophiuchus, and exhibits an abundance of sources with high gas column surface density for a given YSO density. Sources in NGC 2264 follow a different relationship than those in Mon OB1 cloud, exhibiting a larger YSO surface density for a given gas column density on average, and a steeper power law. The over-density of YSOs in this region of star formation
suggests that the natal gas has been partially dispersed, possibly due to Allen's
source, the massive B star. Outside of NGC 2264 in Mon OB1 cloud, we see an under-dense
population of YSOs compared to the gas column density. The northern portion of Mon OB1 cloud  contains
small groups of star formation ($\sim$few YSOs) which include mostly Class 0/I sources and the southern region of Mon OB1 cloud has few YSOs (see Figure \ref{extinction}). This suggests that star formation may just be
starting in these regions. Overall, the the fact that we see a spread in the $\Sigma_{YSO}$ vs. $\Sigma_{gas}$ relation for both NGC 2264 and Mon OB1 cloud  suggests that Mon OB1 East contains regions of YSOs at different evolutionary stages, just like that observed in the Cep OB3 and Perseus molecular clouds  \citep{Gutermuth2011}.

\begin{figure}
\includegraphics[scale=1]{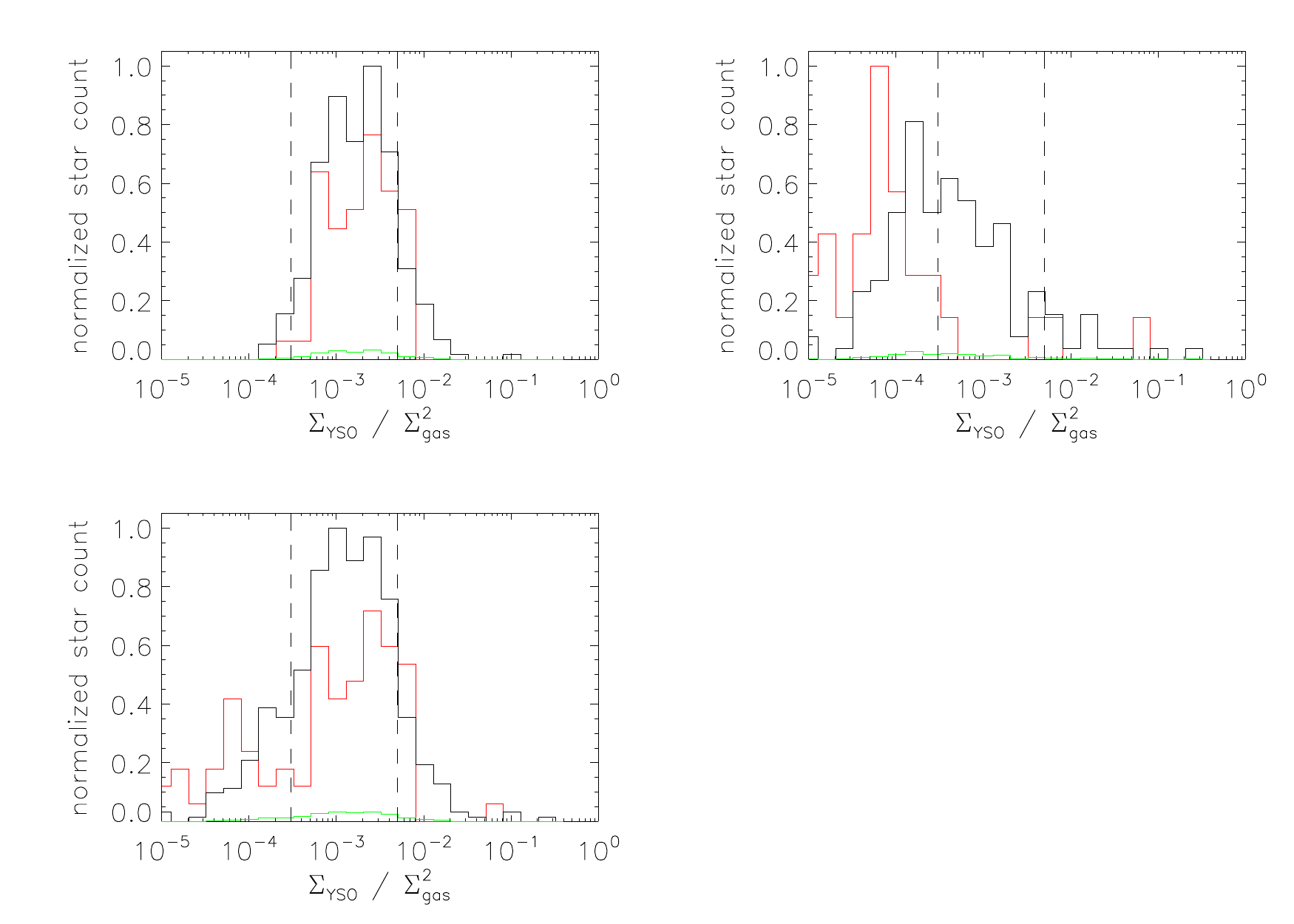}
\caption{\label{ratio} Histograms of $\Sigma_{YSO} / \Sigma^{2}_{gas}$ for NGC 2264 (top left), Mon OB1 cloud (top right), and all of Mon OB1 East (bottom), separated by evolutionary class. Histograms for Class II YSOs are in black, normalized to the peak bin. Class I YSOs are in red, and are multiplied by a factor of 3.7, the median ratio of Class II to Class I sources in nearby clusters \citep{Gutermuth2009} and then normalized to the peak bin. Green is the Class II histogram, normalized to the peak bin and divided by 30. This is the upper limit to the expected frequency of edge on Class II sources that could be misclassified as Class 0/I objects.\citep{Gutermuth2009, Gutermuth2011}. Vertical dashed lines mark the boundary between young protostar rich regions (left), embedded cloud populations (center) and regions where gas has been dispersed \citep[right;][]{Gutermuth2011}  } 
\end{figure}

To further explain the spread in the data and the idea that Mon OB1 East  contains regions of differing evolutionary stages, we construct histograms of $\Sigma_{*} / \Sigma^{2}_{gas}$ for all Class 0/I and Class II sources in Mon OB1 East (Figure \ref{ratio}). \citet{Gutermuth2011} use this technique to discern whether nearby GMCs are coeval in forming stars, or if the clouds exhibit regions of differing evolutionary stages. Figure \ref{ratio} shows the histograms of Class 0/I sources in red, and Class II sources in black, normalized to the peak bin value. The green histogram depicts the Class II sources, normalized to the Class 0/I peak bin multiplied by 3.7  \citep[the median ratio of Class II to Class 0/I sources in nearby clusters;][]{Gutermuth2009}, and divided by 30. This represents the maximum number of sources that could be edge-on Class II YSOs misclassified as Class 0/I YSOs \citep{Gutermuth2009}. The two vertical dashed lines at $\Sigma_{*} / \Sigma^{2}_{gas} = 3\times10^{-4}$ and $5\times10^{-3}$ pc$^{2}$ M$^{-1}_{\odot}$ are derived from the borders of the power law fits to Mon R2 and Ophiuchus and represent the boundary between coeval cloud evolution and more heterogeneous cloud evolution \citep{Gutermuth2011}. Sources with $\Sigma_{*} / \Sigma^{2}_{gas} < 3\times10^{-4}$ and a high fraction of Class 0/I sources indicate regions with an over density of gas  where the onset of star formation is more recent. Sources with $\Sigma_{*} / \Sigma^{2}_{gas} > 5\times10^{-3}$ indicate an over density of stars and thus a later evolutionary stage where gas has either been converted into stellar mass, or dispersed via winds or ionizing radiation. The region between these two extremes represents a stage of ongoing star formation similar to that of clouds evolving coevally. Mon OB1 East shows that most the YSOs fall in the mid-range of $\Sigma_{*} / \Sigma^{2}_{gas}$, but there are significant sources at both extremes. Mon OB1 cloud shows an over density of gas, suggesting that star formation is just beginning in these regions, whereas NGC 2264 is between the two extremes and thus exhibits ongoing star formation, as well as some gas dispersal. Thus, Mon OB1 East is similar to other heterogeneously evolving clouds, of which Cep OB3 and Perseus are the best examples in \citep{Gutermuth2011}

Following the methods described in \citet{Myers2012}, we can use the protostar fraction (i.e. the ratio of Class 0/I YSOs to all YSOs) to determine the approximate age of Mon OB1 East. We find a total of 105 class 0/I YSOs and 1118 Class II, III and transition disk sources (when considering all \emph{Spitzer} classified disk sources) in Mon OB1 East , resulting in a protostar fraction of 0.086. Assuming a constant protostar birthrate and equally likely accretion stopping, use of equations 36-39 in \citet{Myers2012} result in a star-forming age of $\sim$2 Myr, and a mean birthrate of 612 stars/Myr. These results are similar to those for the GMC RCW 38  \citep{Myers2012}. Based on our calculations, Mon OB1 East is considerably younger than the previous age estimate for NGC 2264 of $\sim$5 Myr \citep{Dahm2005}. This is not surprising as we are including regions of Mon OB1 East that appear less evolved in our age estimate, not just those sources in NGC 2264. Furthermore, \citet{Dahm2005} used visible light photometry and spectroscopy and were not able to detect younger, more embedded sources in NGC 2264 itself.

Overall, Mon OB1 East appears very similar to other molecular clouds such as Perseus, Ophiuchus, Cep OB3b and Orion in that there is distinct clustered regions of star formation, associated with lower density regions of YSOs. There is strong evidence for ongoing star formation in NGC 2264, as well as in small groups in the North and South-East of Mon OB1 cloud. Thus, it is peculiar that the number of Class III sources towards Mon OB1 cloud is larger than towards NGC 2264, resulting in a small disk fraction. This may be due to an older, more dispersed population of stars associated with the cloud which no longer shows bright IR-excesses; such a population may be the result of a previous generation of star formation.  This more dispersed population would not strongly alter the disk fraction of the NGC 2264, where the high density of younger stars is much higher than that of the more dispersed population.  In contrast, for the spatially more extended  Mon OB1 cloud, the number of stars in the dispersed population can exceed the small number of dusty YSOs identified by Spitzer resulting in a low disk fraction.  Future visible light and X-ray observations are needed to identify older pre-main sequence stars toward NGC 2264, confirm the existence of this dispersed population, and ascertain whether these stars are slightly in the foreground, yet associated with the cloud, or currently embedded within the cloud.

\section{Summary}

We presented Spitzer aperture photometry for 10,454 YSOs, contaminants, and field stars towards Mon OB1 East giant molecular cloud. This includes the NGC 2264 cluster as well as the more extended cloud region which we refer to as Mon OB1 cloud. We classified YSOs via their infrared colors using the method developed by \citet{Gutermuth2008,Gutermuth2009}, and found 584 total disk sources, 105 of which are Class O/I protostars. The dusty YSOs with IR-excesses are densely clustered in the NGC 2264 regions, but also are found at a much lower density in the extended cloud region. This suggests that this extended cloud contains both clustered and distributed star formation.  Using K$_{S}$-band magnitude histograms, we statistically estimated the number of field stars in the line of sight, and thus the number of Class III pure-photosphere sources within Mon OB1 East. We found a resulting disk fraction of 45 $\pm$4\% within NGC 2264 and 19 $\pm$5\% for the Mon OB1 cloud. 

The NGC 2264 cluster shows a correlation between YSO surface density and molecular gas column density similar to that found in the Mon R2 molecular cloud, but exhibits considerable spread in the data. In contrast, the Mon OB1 Cloud shows a higher gas column density for a given YSO surface density as compared to Mon R2, Ophiuchus and the NGC 2264 cluster.  Furthermore, Mon OB1 cloud is rich in protostars, suggesting that it may be undergoing a recent episode of star formation. These data are evidence for heterogenous star formation in Mon OB1 East, with the current star formation in the Mon OB1 cloud region being more recent than that in the NGC 2264 cluster, and an older more dispersed population of stars likely being responsible for the large number of Class III objects and the low disk fraction in the Mon OB1 cloud region.

\section*{Acknowledgements}
We gratefully acknowledge funding support from NASA ADAP grants NNX11AD14G and NNX13AF08G and Caltech/JPL awards 1373081, 1424329, and 1440160 in support of Spitzer Space Telescope observing programs. This work received support through that provided to the IRAC and MIPS instruments by NASA through contracts 960541 and 960785, respectively, issued by JPL. Support for this work was also provided by NASA through awards to STM and JLP issued by JPL/Caltech 

We would also like to thank Bo Reipurth for providing the $^{13}$CO map on Mon OB1 East. This publication makes use of data products from the Two Micron All Sky Survey, which is a joint project of the University of
Massachusetts and the Infrared Processing and Analysis Center/California
Institute of Technology, funded by the National Aeronautics and Space
Administration and the National Science Foundation. This research has made use of the VizieR
catalogue access tool, CDS, Strasbourg, France. This work is based on
observations made with the Spitzer Space Telescope, which is operated by the
Jet Propulsion Laboratory, California Institute of Technology under a contract
1407 with NASA.

\appendix
\section*{APPENDIX}
\section{Uncertainties in Disk fraction calculations}
The error in disk fraction is determined by adding in quadrature the uncertainties due to background contamination corrections and the uncertainty in the disk fraction calculated via Poisson statistics. We can equivalently express the disk fraction calculated using KHM histograms as
\begin{equation}
f_{disk} = \frac{N_{disk}}{N_{member}} = \frac{N_{disk}}{N_{on} - \frac{\Omega_{on}}{\Omega_{off}}N_{off}}
\end{equation}
where N$_{disk}$ is the number of sources with disks towards the region of interest, N$_{on}$ is the total number of sources (all disk sources $+$ Class III/field stars) towards the region of interest, N$_{off}$ is the total number of sources towards the reference region, and $\frac{\Omega_{on}}{\Omega_{off}}$ is the ratio of the solid angle subtended by the region of interest and the reference region. The main uncertainty in our disk fraction calculation is the uncertainty in the number of contaminating background stars, and can be expressed as
\begin{equation}
\sigma(N_{member}) = \sqrt{\frac{\Omega_{on}}{\Omega_{off}} N_{off} + \left(\frac{\Omega_{on}}{\Omega_{off}}\right)^{2}N_{off}}.
\end{equation}
The first term is the Poisson noise in the number of field stars towards the reference region, and the second is the uncertainty in our count of the number of field stars. Thus, the uncertainty in the ratio of YSOs with disks  to total YSOs in the region of interest is
\begin{equation}
\sigma(N_{disk}/N_{member}) = \frac{N_{disk}}{(N_{on} - \frac{\Omega_{on}}{\Omega_{off}}N_{off})^2}\ \sqrt{\frac{\Omega_{on}}{\Omega_{off}} N_{off} + \left(\frac{\Omega_{on}}{\Omega_{off}}\right)^{2}N_{off}}.
\end{equation}
If we assume that the disappearance of the gas rich primordial and transition disks detected by Spitzer is a Poisson process, than the uncertainty in the disk fraction is given by
\begin{equation}
\sigma(f_{disk,Poisson}) = \frac{\sqrt{N_{disk}}}{N_{member}}.
\end{equation}
Adding the Poisson disk fraction uncertainty in quadrature with the uncertainty in the ratio of YSOs with disks  to total YSOs, we obtain the total uncertainty in the disk fraction:
\begin{equation}
\sigma(f_{disk}) = \left(\frac{N_{disk}}{(N_{on} - \frac{\Omega_{on}}{\Omega_{off}}N_{off})^2} +  \frac{N^2_{disk}}{(N_{on} - \frac{\Omega_{on}}{\Omega_{off}}N_{off})^4}\left(\frac{\Omega_{on}}{\Omega_{off}}\right)\ N_{off} + \frac{N^2_{disk}}{(N_{on} - \frac{\Omega_{on}}{\Omega_{off}}N_{off})^4}\left(\frac{\Omega_{on}}{\Omega_{off}}\right)^2\ N_{off}\right)^{1/2}.
\end{equation}

\bibliography{val_refs}
\pagebreak


\begin{deluxetable}{ccccccccccc}
\tabletypesize{\scriptsize}
\tablecaption{Catalog of sources detected with \emph{Spitzer} Space Telescope IRAC
and MIPS instruments which are classified as potential YSOs/field stars\tablenotemark{1} }
\tablewidth{0pt}

\tablehead{Source & $\alpha_{J2000}$ & $\delta_{J2000}$ & $[3.6]$  & $[4.5]$  & $[5.8]$  & $[8.0]$  & $[24.0]$ & Class\tablenotemark{2}  & S08/09 Class\tablenotemark{3} }

\startdata
14 & 06 43 14.76 & 08 44 58.38 & 12.78 $\pm$ 0.007 & 13.07 $\pm$ 0.009 & ........ & 13.21 $\pm$ 0.132 & ........ & III/F & ........\\ 
17 & 06 43 09.64 & 08 45 00.18 & 14.07 $\pm$ 0.022 & 14.02 $\pm$ 0.022 & ........ & ........ & ........ & III/F & ........\\ 
21 & 06 43 13.37 & 08 45 02.15 & 12.47 $\pm$ 0.005 & 12.40 $\pm$ 0.006 & ........ & 12.45 $\pm$ 0.055 & ........ & III/F & ........\\ 
26 & 06 42 41.52 & 08 45 06.44 & 13.78 $\pm$ 0.011 & 13.65 $\pm$ 0.015 & 13.68 $\pm$ 0.105 & ........ & ........ & III/F & ........\\ 
34 & 06 43 10.32 & 08 45 11.33 & 10.32 $\pm$ 0.003 & 10.17 $\pm$ 0.003 & 10.07 $\pm$ 0.005 & 10.03 $\pm$ 0.008 & ........ & III/F & ........\\ 
36 & 06 43 03.57 & 08 45 12.55 & 14.37 $\pm$ 0.013 & 14.27 $\pm$ 0.023 & ........ & ........ & ........ & III/F & ........\\ 
37 & 06 42 48.26 & 08 45 13.24 & 13.94 $\pm$ 0.009 & 13.90 $\pm$ 0.015 & 13.72 $\pm$ 0.091 & ........ & ........ & III/F & ........\\ 
45 & 06 42 42.63 & 08 45 20.25 & 11.62 $\pm$ 0.003 & 11.60 $\pm$ 0.004 & 11.55 $\pm$ 0.016 & 11.56 $\pm$ 0.035 & ........ & III/F & ........\\ 
50 & 06 42 52.48 & 08 45 23.60 & 13.94 $\pm$  0.009 & 13.88 $\pm$  0.013 & 13.91 $\pm$  0.089 & ........ & ........ & III/F & ........ \\
54 & 06 43 02.63 & 08 45 23.95 & 13.38 $\pm$  0.007 & 13.28 $\pm$  0.011 & 13.27 $\pm$  0.050 & 13.78 $\pm$  0.155 & ........ & III/F & ........ \\
\enddata

\tablenotetext{1}{A complete version of this table is available in the online version}  
\tablenotetext{2}{Classification: 0/I: Class 0/I, II: Class II, TD: Transition disk, III/F: Class III/field star, AGN: active galactic nuclei, PAH: polycyclic aromatic hydrocarbon emitting source, SHOCK: shocked emission  gas source.}
\tablenotetext{3}{Classification: none: \emph{Spitzer} detected source not part of the cluster. Other classifications explained in \citet{Sung2008, Sung2009}}

\end{deluxetable}


\begin{deluxetable}{ccccccccccc}
\tablecaption{YSO counts and disk fractions for Mon OB1 East}
\tablewidth{0pt}
\tablehead{Region  & Excess sources\tablenotemark{1}  & Class III sources  & Disk fraction}

\startdata
NGC 2264 & 220  & 264  & 45 $\pm 4\%$\\

Mon OB1 cloud & 88  & 375  & 19 $\pm 5\%$\\
\enddata
\tablenotetext{1}{These are the the number of \emph{Spitzer} identified excess sources that satisfy our dereddened J-band magnitude, extinction, K$_{S}$-band magnitude criteria. }  
\end{deluxetable}


\end{document}